\documentstyle[12pt,epsfig]{article}
\newcommand\slv{v\kern-5pt\raise1pt\hbox{$\scriptstyle/$}\kern1pt}

\begin{document}
\thispagestyle{empty}
\begin{flushright}
WUE/ITP-97-015\\
MPI-PhT/97-34\\
SPhT-T97/042
\end{flushright}
\vspace{0.5cm}
\begin{center}
{\Large \bf Perturbative QCD Correction }\\[.3cm]
{\Large \bf to the $B\to\pi$ Transition Form Factor}\\
\vspace{1.7cm}
{\sc \bf A.~Khodjamirian$^{1,a} $, R.~R\"uckl$^{1,2}$, S.~Weinzierl$^3$, O.~Yakovlev$^{1,b}$}\\[1cm]
\begin{center} \em $^1$ Institut f\"ur Theoretische Physik, Universit\"at W\"urzburg,
D-97074 W\"urzburg, Germany \\
\vspace{4mm}
$^2$ Max-Planck-Institut f\"ur Physik, 
Werner-Heisenberg-Institut, D-80805 M\"unchen, Germany \\
\vspace{4mm}
$^3$ Service de Physique Th\'eorique, Centre d'Etudes de Saclay,\\ 
F-91191 Gif-sur-Yvette Cedex, France
\end{center}\end{center}
\vspace{2cm}
\begin{abstract}\noindent
{We report on the perturbative $O(\alpha_s)$ correction 
to the light-cone QCD sum rule for the  
$B\to \pi$ transition form factor $f^+$.
The correction to the product $f_Bf^+$ in leading twist approximation
is found to be about 30\%, that is similar in 
magnitude to the corresponding $O(\alpha_s)$ correction in 
the two-point sum rule for $f_B$. The resulting   
cancellation of large QCD corrections in $f^+$  
eliminates one important uncertainty in the sum-rule 
prediction for this form factor. 
}    
\end{abstract}

\vspace*{\fill}

\noindent $^a${\small \it On leave from 
Yerevan Physics Institute, 375036 Yerevan, Armenia } \\
$^b${\small \it On leave from Budker Institute of Nuclear Physics (BINP),
630090, Novosibirsk, Russia }
\newpage

%\section {Introduction}

{\bf 1.}   The semileptonic decay $B\to\pi l\nu_l $ 
is one of the most important reactions for the 
determination of the CKM parameter 
$V_{ub}$. However, in order to extract $V_{ub}$  
from data one needs an accurate 
theoretical calculation of    
 the hadronic matrix element
\begin{eqnarray}\label{formdef}
\langle\pi (q)|\bar u\gamma_{\mu}b |B(p+q)\rangle
=2f^+(p^2)q_{\mu}+(f^+(p^2)+
f^-(p^2))p_{\mu},
\end {eqnarray}
where $p+q$, $q$ and $p$ denote the $B$ and $\pi$ four-momenta 
and the momentum transfer, respectively, and $f^{\pm}$ are two 
independent form factors.  

A very reliable approach 
to calculate $f^{\pm}$ in the framework of QCD  
is provided by the operator
product expansion (OPE) on the light-cone \cite{CZ,ER,BL} in 
combination with QCD sum rule techniques.
The sum rule for the form factor $f^{+}(p^2)$ has been 
obtained in \cite{BKR,BBKR} taking into account all 
twist 2, 3 and 4 operators, while $f^-(p^2)$ is derived 
in \cite{KRW}.    
The most important missing elements of these calculations 
are the perturbative QCD corrections to the correlation function 
leading to (\ref{formdef}). Here we report on a calculation 
of the $O(\alpha_s)$ correction  
to $f^+$ which eliminates one of the main 
uncertainties in the existing sum rule results.

The calculation  has several aspects which are worth pointing out.
Firstly, the sum rule is actually derived for the product 
$f_Bf^+$, $f_B$ being  
the $B$ meson decay constant defined by
\begin{equation}
\label{fBB} 
\langle B|\bar bi\gamma_5 d|0 \rangle =m^2_Bf_B/m_b~.
\end{equation}
The form factor $f^+$ itself is then obtained by dividing out 
$f_B$ taking  the value determined from  
the corresponding two-point QCD sum rule. 
In previous estimates, 
the $O(\alpha_s)$ correction to $f_B$ 
was thereby ignored for consistency because of the lack of the 
$O(\alpha_s)$ correction to $f_Bf^+$. Our calculation 
now allows to take 
into account the correction to $f_B$ which
is known to be sizeable.  
Secondly, knowing the $O(\alpha_s)$ corrections, also 
the heavy quark mass entering the sum rule can be 
properly defined.
Thirdly, perturbative corrections to exclusive amplitudes 
involving light-cone wave functions have so far been studied 
only for massless quarks. For example, in \cite{Chase,Braaten,kad} 
the amplitude of the pion transition to
two virtual photons was calculated to $O(\alpha_s)$. 
The calculation for a finite quark 
mass is new and will have numerous applications.

The main result of our work is the following. 
The $O(\alpha_s)$ correction 
to the light-cone sum rule for the product $f_Bf^+$ 
calculated in the leading twist approximation 
is about 30\%  and positive. 
Since the $O(\alpha_s)$ correction to $f_B$ is similar in size
and of the same sign, the large 
QCD corrections cancel in $f^+$ making the prediction 
of the form factor very reliable, at least from the point of view of 
perturbative QCD.   

In this letter, we outline our
calculation, present the final analytical results,
and give first numerical estimates. Technical details,
a thorough numerical analysis,    
and further applications will be presented elsewhere.
\bigskip

{\bf 2.} The light-cone sum rule 
for the form factor $f^+$ was derived in \cite{BKR}. We repeat 
here the necessary points. In QCD, the correlation
function of two heavy-light currents,
 \begin{eqnarray}
 F_{\mu}(p,q)&=&i\int dxe^{ip\cdot x}
 \langle\pi (q)|T\{\bar u (x)\gamma_\mu b(x) , m_b\bar b(0)i\gamma_5 
 d(0)\}|0\rangle
\nonumber
\\
 &=& F(p^2,(p+q)^2)q_\mu+\tilde F(p^2,(p+q)^2)p_\mu~,
\label{corr} 
\end{eqnarray}
can be calculated in the region $(p+q)^2<0 $ and  
$p^2 < m_b^2 - O($1GeV$^2)$ using OPE near the light-cone, 
i.e. at $x^2 \simeq 0$. 
In (\ref{corr}), we have multiplied the pseudoscalar current by the 
$b$-quark mass in order to assure renormalization-group invariance 
of the correlation function.
After contracting the $b$-quark 
fields in (\ref{corr}), $F_\mu$ 
is expressed in terms of bilocal matrix elements of increasing twist.
In the present calculation we focus on 
the leading twist 2 contribution 
which enters through the following matrix element:
\begin{eqnarray} 
\left\langle \pi(q) \left|
 \bar{u}(x) \gamma_{\mu} \gamma_{5}
 P \exp \left( i g_s \int\limits_{0}^{1} d\alpha ~x \cdot A(
\alpha x) \right)d(0)
\right| 0 \right\rangle=
- i q_{\mu} f_{\pi} \int\limits_{0}^{1} du \varphi_\pi(u) e^{i u q \cdot x} 
+ ...~,
\label{wave}
\end{eqnarray}
where the ellipses stand for terms of higher twist. 
The path-ordered gluon operator ensures gauge invariance.
In the light-cone gauge, $x\cdot A = 0$, 
adopted here as usual  this operator is unity.
The distribution function $\varphi_\pi(u)$ is known as the 
twist 2 light-cone wave function of the pion \cite{CZ,ER,BL}.

Comparison of (\ref{formdef}) and (\ref{corr}) shows that in
order to calculate $f^+$ one only has to deal with the invariant 
amplitude $F$ in (\ref{corr}). With (\ref{wave}), $F$ can be 
written as a convolution 
of a hard amplitude $T(p^2,(p+q)^2,u)$ 
calculable within  perturbation
theory, with the pion wave function  
$\varphi_{\pi}(u)$ containing the long-distance effects:
\begin{eqnarray}\label{represent}
F(p^2,(p+q)^2)=-f_\pi\int^1_0 du \varphi_\pi (u) T(p^2,(p+q)^2,u).
\end{eqnarray}
In zeroth order in $\alpha_s$, the hard amplitude represented 
graphically in Fig. 1a reads 
\begin{eqnarray}\label{Born}
 T_0(p^2,(p+q)^2,u)=\frac{m_b^2}{p^2(1-u)+(p+q)^2u-m_b^2}.
\end{eqnarray}
At fixed $p^2 <m_b^2$, 
$F$ is an analytic function in the complex 
$(p+q)^2 $-plane, with a cut along the real axis starting from 
$(p+q)^2 =m_b^2$.

One can therefore write a dispersion relation
\begin{eqnarray}
F(p^2,(p+q)^2)=\int^{\infty}_{m_b^2}\frac{\rho^{QCD} (p^2,s)ds}
{s-(p+q)^2}~.
\label{QCD}
\end{eqnarray}
Equating the 
QCD result obtained  
with $\rho^{QCD}= \frac{1}{\pi} \mbox{Im} F$ and
the hadronic representation of $F$ 
following from (\ref{QCD}) with the spectral density 
\begin{eqnarray}
\rho(p^2,s)=\delta (s-m^2_B)2m^2_Bf_Bf^+(p^2)
+\rho^{QCD}(p^2,s)\Theta (s-s_0)
\label{spectr}
\end{eqnarray}
yields the desired relation between $f^+$ and the invariant 
function $F$. In (\ref{spectr}), the first term stems from 
the $B$ ground state, whereas the second term represents the 
contributions from the higher
resonances and the continuum in the $B$--meson channel 
above the threshold $s_0$. 
Invoking quark-hadron duality the latter is replaced 
by the spectral density $\rho^{QCD}$. 
The sum rule finally follows from the above
after Borel transformation in $(p+q)^2$:
\begin{eqnarray}
f_Bf^+(p^2)=\frac{1}{2m^2_B}\int^{s_0}_{m_b^2} \rho^{QCD}(p^2,s)
e^{\frac{m^2_B-s}{M^2}}ds~,
\label{sr}
\end{eqnarray}
where
\begin{eqnarray}
\rho^{QCD}(p^2,s) 
 = - \frac{f_{\pi}}{\pi}  \int\limits_0^1 du \varphi_\pi (u) 
\mbox{Im} T(p^2,s,u)~.
\label{rho}
\end{eqnarray}

With the zeroth order 
approximation (\ref{Born}), one easily obtains 
\begin{equation}
\mbox{Im} T_0(p^2,s,u)= -\pi\delta(1-\frac{p^2}{m_b^2}(1-u) 
-\frac{s}{m_b^2}u ).
\label{imT0}
\end{equation}
Substitution of (\ref{rho}) and (\ref{imT0}) 
in (\ref{sr}) and integration over $s$ reproduce the leading twist 2 
contribution to the light-cone sum rule given in \cite{BKR}.
In this approximation the evolution of $\varphi_\pi$ is 
taken into account in the leading order (LO). In order to go to 
the next-to-leading order (NLO), one has to calculate the $O(\alpha_s)$
correction to $\mbox{Im}T$ and use the NLO-evolution of $\varphi_\pi$.
This problem is solved below.

%%%%%%%%%%%%%%%%%%%%%%%%%%%%%%%%%%%%
\bigskip

{\bf 3.} The first step is to calculate the 
$O(\alpha_s)$ correction to the hard amplitude $T$ 
which we write as
\begin{eqnarray}
T(r_1,r_2,u)=T_0(r_1,r_2,u)+\frac{\alpha_sC_F}{4\pi}T_1(r_1,r_2,u)~,
\label{t1}
\label{T01}
\end{eqnarray}
introducing convenient dimensionless 
variables $r_1 = p^2/m_b^2$
and $r_2=(p+q)^2/m_b^2$. The zeroth order amplitude 
$T_0$ is given in ({\ref{Born}). In Figs. 1b - 1g  
we show the Feynman diagrams determining the first order amplitude 
$T_1$.
The calculation is performed 
in general covariant gauge in order 
to have a possibility to check the gauge invariance of the result. 
Both the ultraviolet (UV)  and infrared
divergences are regularized by 
dimensional regularization and renormalized 
in the $\overline{MS}$ scheme 
with  totally anticommuting $\gamma_5$.  
This choice is motivated by the fact that the same 
scheme is used in the 
calculation of the NLO evolution kernel of the wave function 
$\varphi_\pi(u)$ \cite{Rad1}. 

From the diagrams depicted in Fig. 1 we find 
\begin{eqnarray}\label{resultT1}
&T_1(r_1,r_2,u)&=
  \frac{3(1+\rho)}{(1-\rho)^2} 
 \left( 
            \Delta - \ln\frac{m_b^2}{\mu^2} 
                                                                + 1 \right)
       -\frac{2}{1-\rho} \left[ 2 \tilde{G}\left(\rho\right) -
 \tilde{G}\left(r_1\right) - \tilde{G}\left(r_2\right) \right] 
\nonumber \\
 & &      +\frac{2}{(r_1-r_2)^2} \left(
             \frac{1-r_2}{u} \left[ \tilde{G}\left(\rho\right) - 
\tilde{G}\left(r_1\right)\right]
           + \frac{1-r_1}{1-u} \left[ \tilde{G}\left(\rho\right) -
 \tilde{G}\left(r_2\right)\right] \right)  \nonumber\\
 & &      +\frac{\rho+(1-\rho)\ln\left(1-\rho\right)}{\rho^2}
       -\frac{2}{1-\rho} \frac{(1-r_2)\ln\left(1-r_2\right)}{r_2}
+\frac{3-\rho}{\left(1-\rho\right)^2}
\nonumber \\
 & &      -\frac{2}{(1-u)(r_1-r_2)}
             \left( \frac{ (1-\rho) \ln\left(1-\rho\right)}{\rho}
                  - \frac{ (1-r_2) \ln\left(1-r_2\right)}{r_2}  \right) 
\end{eqnarray}
with   
\begin{eqnarray}
&&\Delta = \frac{2}{4-d}-\gamma_E+\ln (4\pi ),\quad \quad 
\rho = r_1 + u (r_2-r_1), \\
\tilde{G}\left(\rho \right) 
& = & \mbox{Li}_2(\rho) + \ln^2(1-\rho) -\ln(1-\rho)
\left(\Delta -\ln\frac{m_b^2}{\mu^2}
 +1 \right),\nonumber 
\end{eqnarray}
$\mbox{Li}_2(x)=-\int\limits^x_0\frac{dt}t \ln(1-t)$ being the Spence function.
The UV renormalization scale 
and the factorization
scale of the collinear (COL) divergences  are taken  
to be equal and denoted by $\mu$. 
In order to trace the origin of the various divergent
terms we have performed additional explicit calculations.
In particular, we have used mass regularization 
by giving the light quarks a small but finite mass, 
and momentum regularization keeping the light quarks off mass shell.
In this way, we have unambiguously separated the  
COL-divergent terms from  the UV-divergent terms.
The latter add up to
\begin{eqnarray}\label{UVterm}
T^{UV}_1(r_1,r_2,u)= 
\frac{6\rho }{(1-\rho)^2}\Delta.
\end{eqnarray} 

The correlation function (\ref{corr}) involves the unrenormalized 
quark currents $J_{5} = \bar{b} \gamma_5 d$ and  
$J_{\mu} = \bar{u} \gamma_{\mu} b$ as well
as the bare $b$-quark mass $m_b$. As usual, 
we define the corresponding renormalized quantities
by  
$$
J_{5} \to Z_{5} J_{5},\quad
J_{\mu} \to Z_{V} J_{\mu},\quad 
m_b \to Z_m  \hat{m}_b~.
$$
In the $\overline{MS}$-scheme, the renormalization constants are
given by 
\begin{eqnarray}
Z_{5}  =  1+3\Delta\frac{\alpha_SC_F}{4\pi},\quad
Z_{V}  =  1, \quad
Z_{m}  =  1-3\Delta\frac{\alpha_SC_F}{4\pi}~.
\label{z}
\end{eqnarray}
We see that the overall renormalization factor of (\ref{corr}) 
is $Z_mZ_5Z_V=1$. 
The UV-renormalized hard amplitude $T$  then follows from the   
unrenormalized result (\ref{t1})
just by reexpressing  
the unrenormalized mass $m_b$ through the renormalized mass $\hat{m}_b$ .
As a result, an additional $O(\alpha_s)$ 
contribution to $T$ emerges 
which exactly cancels the term $T_1^{UV}$ given in (\ref{UVterm}).

In addition to the UV-divergent terms, the function $T_1$
contains the COL-divergent terms: 
\begin{eqnarray}\label{IR}
T^{COL}_1(r_1,r_2,u)=-\Delta T_0(u)
\Bigg[ 3
  -2\ln\left( \frac{1-r_2}{1-r_1}\right)\frac{(1-r_1)(1-r_2)-
   u(1-r_1)(r_2-r_1)}{u(r_2-r_1)^2} \\
    -2\ln\left( \frac{1-\rho}{1-r_2}\right)
     \frac{(1-r_1)(1-r_2)
      -u(1-u)(r_2-r_1)^2}{u(1-u)(r_2-r_1)^2} \Bigg]~.
\nonumber
\end{eqnarray} 
It is straightforward to 
check that $T^{COL}_1$  can be written in the form
\begin{eqnarray}\label{infraredform}
T^{COL}_1(r_1,r_2,u)&=&-\Delta\frac{1}{2}\int^1_0dwV(w,u)T_0(r_1,r_2,w)~, 
\end{eqnarray} 
where $V(w,u)$ is the kernel of the Brodsky-Lepage 
evolution equation \cite{BL} 
of the light-cone wave function $\varphi_\pi(u)$ 
introduced in (\ref{wave}):
\begin{equation}
d\varphi_\pi (u,\mu)/d\ln \mu = \int^1_0 d\omega V(u,\omega)
\varphi_\pi(\omega,\mu) 
\label{BLL}
\end{equation}
with
\begin{eqnarray} 
V(w,u)  = \frac{\alpha_s(\mu) C_{F}}{\pi} \left[ \frac{1-w}{1-u} 
\left( 1 + \frac{1}{w-u} \right) \Theta(w-u)
+ \frac{w}{u} \left( 1 + \frac{1}{u-w} \right)
 \Theta(u-w) \right]_{+}~.
\label{BL}
\end{eqnarray}
The operation $+$ is defined by 
\begin{eqnarray}
 V(w,u)_{+} & = & V(w,u) - \delta(w-u) \int\limits_{0}^{1} V(v,u) dv. 
\label{plus}
\end{eqnarray}
The appearance of COL-divergent terms 
in the hard amplitude $T$ in the form $(\ref{infraredform})$
reflects the factorization of 
the correlation function  into a wave function and a 
hard amplitude \cite{ER,Braaten,kad}. 
For the  factorization scheme we have  
again adopted the $\overline{MS}$-scheme, 
i.e. we have subtracted the terms in the UV-renormalized 
hard scattering amplitude proportional to $\Delta$.
These are the terms absorbed in the definition 
of the scale-dependent wave function. 
The remaining $\mu$-dependences 
of the hard scattering amplitude and of the wave function
compensate each other. 

Up to now we have worked in the $\overline{MS}$-scheme.
However, the $\overline{MS}$ quark mass depends explicitly on the 
renormalization scale $\mu$ and implicitly on the renormalization
prescription. A renormalization-scheme-independent definition of the 
quark mass within QCD perturbation theory
is given by the pole mass which we denote $m_b^*$. 
Since we intend to use the set of parameters $(m_b^*,
 f_B, s_0)$ determined self-consistently
from an independent analysis of the two-point sum rule 
for $f_B$ \cite{fB}
it is  convenient to replace $\hat{m}_b$ by $m_b^*$ 
also in the sum rule for $f_Bf^+$ 
using the well-known one-loop relation: 
\begin{eqnarray}
\hat{m}_b& = & m_b^* 
\left( 1 + \frac{\alpha_S C_F}{4 \pi} \left(-4 
         + 3 \ln \frac{m_b^{*2}}{\mu^2} \right) \right)~.
\label{mass}
\end{eqnarray}
To $O(\alpha_s)$, this replacement adds the term 
\begin{eqnarray}
- \frac{2\rho}{(1-\rho)^2}  \left( 
       4 - 3\ln\frac{m_b^{*2}}{\mu^2} \right)
\end{eqnarray}
to the renormalized amplitude 
$T_1$. The  final result for the invariant function (\ref{represent}) 
then reads 
\begin{eqnarray}\label{Ffinal}
F(r_1,r_2)=
-f_\pi\int\limits^1_0du\varphi_\pi(u,\mu)T(r_1,r_2,u,\mu)~,
\end{eqnarray}
where the renormalized 
hard amplitude is given by  
$$
T(r_1,r_2,u,\mu) =  
\frac1{\rho-1}
+\frac{\alpha_s(\mu)C_F}{4\pi}
 \Bigg\{\frac1{\rho-1}( -4+3
\ln \frac{m_b^{*2}}{\mu^2}) 
+\frac{2}{\rho-1} \left[ 
     2 G\left(\rho\right) - G\left(r_1\right) - G\left(r_2\right)
 \right] 
$$
%\nonumber
% \\
$$       +\frac{2}{(r_1-r_2)^2} \left(
             \frac{1-r_2}{u} \left[ G\left(\rho\right) -
 G\left(r_1\right)\right]
           + \frac{1-r_1}{1-u} \left[ G\left(\rho\right) - 
G\left(r_2\right)\right] \right) 
$$
%\nonumber\\
$$
       +\frac{\rho+(1-\rho)\ln\left(1-\rho\right)}{\rho^2}
+\frac{2}{\rho-1} \frac{(1-r_2)\ln\left(1-r_2\right)}{r_2} 
           -\frac{2}{\rho-1}
$$
%\nonumber\\
\begin{equation}
       -\frac{2}{(1-u)(r_1-r_2)}
             \left( \frac{ (1-\rho) \ln\left(1-\rho\right)}{\rho}
                  - \frac{ (1-r_2) \ln\left(1-r_2\right)}{r_2}  \right)
    \Bigg\} ~.
\label{result}
\end{equation}
Here 
$G(\rho) = \tilde{G}(\rho )|_{\Delta =0}$, and $\varphi_\pi(u,\mu)$ is 
the pion wave function evolved to the scale $\mu$ in NLO.

To proceed further according to 
(\ref{QCD}) and (\ref{rho}) we calculate  
the imaginary part of the hard scattering amplitude (\ref{result})
for $r_2>1$ and $r_1<1$: 

$$-\frac{1}{\pi} \mbox{Im}T(r_1,r_2,u,\mu) = \delta(1-\rho)
$$
$$
+ \frac{\alpha_s(\mu) C_F}{4 \pi} \left\{
  \delta(1-\rho) \left[ \pi^2 -6 + 3\ln \frac{m_b^{*2}}{\mu^2}   
 -2 \mbox{Li}_2(r_1) \right. \right. 
$$
$$ + 2 \mbox{Li}_2(1-r_2)
\left.  -2 \left( \ln \frac{r_2-1}{1-r_1} \right)^2 + 2 
\left( \ln r_2 +\frac{1-r_2}{r_2} \right)
                     \left( 2 \ln(r_2-1) - \ln(1-r_1) \right)
 \right]$$
$$+ \theta(\rho-1)\left[ 8 \left. \frac{\ln(\rho-1)}{\rho-1}
\right|_+
 +2 \left( \ln r_2 + \frac{1}{r_2} -2 
-2 \ln(r_2-1) + \ln 
\frac{m_b^{*2}}{\mu^2} \right)
             \left. \frac{1}{\rho-1} \right|_+ \right.$$
$$
-2 \frac{r_2-1}{(r_1-r_2)(\rho-r_1)} 
   \left( \ln \rho -2 \ln(\rho-1) + 1 -
\ln \frac{m_b^{*2}}{\mu^2} 
\right)
$$
$$
+2 \frac{1-r_1}{(r_1-r_2)(r_2-\rho)} \left( \ln \frac{\rho}
{r_2} -2 \ln \frac{\rho-1}{r_2-1} \right) 
\left. -4 \frac{\ln \rho}{\rho-1} + 2 \frac{1}{r_2-\rho}
 \left( \frac{1}{\rho} - \frac{1}{r_2} \right)
   +\frac{1}{\rho^2} -\frac{1}{\rho} \right]$$  
$$ 
+\theta(1-\rho) \left[ 2 \left( \ln r_2 + \frac{1}{r_2} 
-2 \ln(r_2-1) 
-\ln \frac{m_b^{*2}}{\mu^2}
\right) \left. \frac{1}{\rho-1} \right|_+
 \right. $$
\begin{eqnarray}
\label{spden}
 -2 \frac{1-r_1}{(r_1-r_2)(r_2-\rho)} \left( \ln r_2 + 1
 -2 \ln(r_2-1) 
-\ln \frac{m_b^{*2}}{\mu^2}
\right) \left. \left. -2 \frac{1}{r_2-\rho} 
\frac{1-r_2}{r_2} \right] \right\}
\end{eqnarray} 
Here, the operation $+$  is defined by 
\begin{eqnarray}
\int d\rho f(\rho) \left. \frac{1}{1-\rho} \right|_+ & = &
  \int d\rho \left( f(\rho) - f(1) \right) \frac{1}{1-\rho}.
\end{eqnarray}
This prescription takes care of the 
spurious infrared divergencies which 
one encounters by taking the imaginary 
part of  (\ref{Ffinal}).
 
Substituting (\ref{spden}) and (\ref{rho}) to 
(\ref{sr}) one obtains the desired sum rule in $O(\alpha_s)$ 
for the form factor $f^+$ in the leading twist 2 approximation:
\begin{eqnarray}
f_Bf^+(p^2)=- \frac{f_{\pi}}{2\pi m^2_B}\int^{s_0}_{m_b^{*2}}ds 
\int\limits_0^1 du ~\varphi_\pi (u,\mu) 
\mbox{Im} T\Big(\frac{p^2}{m_b^{2*}},
\frac{s}{m_b^{*2}},u,\mu\Big)
e^{\frac{m^2_B-s}{M^2}}ds ~.
\label{sumrules}
\end{eqnarray}
The subleading twist 3 and 4 contributions are presently 
known only in zeroth order in $\alpha_s$ \cite{BKR,KRW}.
They will be taken into account in the numerical analysis. 

\bigskip

{\bf 4.} The second step is to determine the decay constant $f_B$ and 
the pion wave function  $\varphi_\pi(u,\mu)$ in NLO. For that purpose  
we have analyzed the two-point sum rule
for $f_B$ obtained from    
the renormalization-group-invariant 
correlation function\\ 
$m_b^2\langle 0\mid T\{J_5^+(x)J_5(0)\} \mid 0 \rangle$   
in $O(\alpha_s)$ \cite{fB}. For the  running coupling constant 
we use the two-loop expression 
with $N_f=4$ and  
$\Lambda^{(4)}=234$ MeV \cite{PDG}
corresponding  to $\alpha_s(M_Z)= 0.112$. 
For $\mu^2$ we take the value   $\mu^2_B= m_B^2-m_b^{*2}$  
corresponding to the average virtuality 
of the correlation function which in turn is given  
by the Borel mass parameter $M^2$. 
With this
choice the following correlated results are extracted from the two-point
sum rule: 
\begin{eqnarray}
\label{fbb}
 f_{B}=180\pm 30\quad\mbox{MeV} \qquad 
m_b^*=4.7\mp0.1\quad\mbox{GeV},\qquad 
s_0=35\pm 2\quad\mbox{GeV}^2.
\end{eqnarray} 
In the following, we adopt the central values in the above intervals.
Note that without $O(\alpha_s)$ correction 
one obtains $f_B = 140 \pm 30 $ MeV.   
The remaining parameters entering (\ref{sumrules}) are 
directly measured: $m_B=5.279 $ GeV and $f_{\pi}=132$ MeV.

The wave function $\varphi_\pi$ 
can be expanded in terms of Gegenbauer
polynomials \\$C_{n}^{3/2}(2u-1)$.
Arguments based on conformal spin expansion
\cite{BF2} allows one to neglect higher terms in this expansion.
We adopt the ansatz suggested in \cite{BF}:
\begin{eqnarray}  
\varphi_\pi(u,\mu_0)=\Psi_0(u)+a_2(\mu_0) \Psi_2(u) 
+a_4(\mu_0) \Psi_4(u),
\end{eqnarray}
where 
$ \quad \Psi_{n}(u) =  6u (1-u) C_{n}^{3/2}(2 u -1)$. 
The asymptotic wave function $\varphi_\pi(u)=6u(1-u)$ is unambigously 
fixed \cite{BL}. The terms $n>0$ describe nonasymptotic 
corrections. The coefficients $a_2(\mu_0)=2/3$ and $a_4(\mu_0)=0.43$
at the scale $\mu_0=500$ MeV 
have been extracted \cite{BF} from a two-point QCD sum rule  
for the moments of $\varphi_\pi(u)$ \cite{CZ}. In NLO, 
the evolution of the wave function 
is given by \cite{kad}:
\begin{eqnarray}
\label{phipi}
\varphi_\pi(u,\mu)= \sum\limits_{n} a_{n}(\mu_{0})
\exp\left(- \int\limits_{\alpha_{s}(\mu_{0})}^{\alpha_{s}(\mu)} d\alpha
 \frac{\gamma^{n}(\alpha)}{\beta(\alpha)} \right)
 \left( \Psi_{n}(u) + \frac{\alpha_{s}(\mu)}{4 \pi}
 \sum\limits_{k>n} d_{n}^{k}(\mu) \Psi_{k}(u) \right)
\end{eqnarray}
with $a_0=1$. The coefficients $d_n^k(\mu )$
are due to mixing effects, induced by the fact that 
the polynomials $\Psi_n(u)$ are
the eigenfunctions of the LO, but not of 
the NLO evolution kernel. The QCD 
beta-function $\beta$ \cite{PDG} 
and the anomalous dimension $\gamma^n$ of the 
$n$-th moment $a_n(\mu)$
of the wave function have to be taken in NLO.
Explicitly \cite{anom},  
\begin{eqnarray}
\gamma^n =\frac{\alpha_s}{4\pi}\gamma_{0}^n
+\left(\frac{\alpha_s}{4\pi}\right)^2\gamma_{1}^n
\label{gamman}
\end{eqnarray}
with 
\begin{eqnarray}
\gamma^{\small 0}_0&=&0, \quad \quad
\quad \gamma^{\small 0}_1=0~,
\nonumber
\\
\gamma^{\small 2}_0&=&\frac{100}{9}, \quad \quad 
\gamma^{\small 2}_1=\frac{34450}{243}-
\frac{830}{81}N_F,
\nonumber
\\
\gamma^{\small 4}_0&=&\frac{728}{45}, \quad  \quad 
\gamma^{\small 4}_1=\frac{662846}{3375}-
\frac{31132}{2025}N_F.
\end{eqnarray}
The NLO mixing coefficients are \cite{kad,Rad1}
\begin{eqnarray}
d_n^k(\mu) & = & \frac{M_{n k}}{\gamma_0^k-\gamma_0^n-2\beta_0}
\left( 1 - \left( \frac{\alpha_s(\mu)}{\alpha_s(\mu_0)} \right)^{\frac{\gamma_0^k-\gamma_0^n-2\beta_0}{2\beta0}}
\right),
\end{eqnarray}
where the numerical values of the first few elements of the matrix 
$M_{n k}$ are 
\begin{eqnarray}
M_{0 2} = -11.2+1.73 N_F,~~ M_{0 4} = -1.41+0.565 N_F, ~~  
M_{2 4} = -22.0+1.65N_F.
\end{eqnarray}
With the above input and (\ref{phipi})
we find $a_2(\mu_B)=0.218$ and $a_4(\mu_B)=0.084$.

Now, we are ready to exploit  
the sum rule (\ref{sumrules}) numerically.
In Fig. 2, the product 
$f_Bf^+(0)$ 
is plotted as a function of the Borel parameter $M^2$.    
The $O(\alpha_s)$ correction turns out to be large, 
between 30\%  and  35\% , and stable 
under variation of $M^2$. More specifically, in the interval 
$M^2= 8 \div 12$ GeV$^2$ we obtain in LO 
\begin{equation}
f_Bf^+(0)=0.0229\div 0.0224\quad\mbox{GeV},
\quad 
f^+(0)=0.163 \div 0.160, 
\end{equation}
and in NLO
\begin{equation}
\label{numbers}
f_Bf^+(0)=0.0306 \div 0.0295 \quad\mbox{GeV},
\quad f^+(0)=0.170\div 0.164
\end{equation}
where $f_B=140$ MeV and 180 MeV 
has been used, respectively. 
Note the almost complete cancellation of the  
NLO correction in $f^+$.  
Furthermore, Fig. 3 shows the momentum 
dependence of the form factor $f^+(p^2)$  
in the region $0<p^2<15\div17$ GeV$^2$
 for $M^2=10$ GeV$^2$, where the sum rule
(\ref{sumrules}) is expected to be valid.
Finally, it is interesting to compare the $\mu$ dependence
in LO and NLO. This is done in Fig. 4. The very
mild $\mu$ -dependence in LO only results from the evolution
of the wave function. In NLO, the $\mu$-dependence is 
stronger than in LO but similar to the $\mu$-dependence 
of $f_B$. As a result, the residual scale dependence 
of $f^+$ is again mild.

The above results refer to the leading twist 2 
approximation. If one adds the LO twist 3 and 
4 contributions, one obtains at $p^2=0$ 
\begin{equation}
f^+(0) = 0.27 ~.
\end{equation}
This value should be compared with the 
LO estimate $f^+(0) = 0.30 $ 
obtained in \cite{BKR,KRW}.

\bigskip
{\bf 5. } 
In this paper, 
we have presented the perturbative QCD correction
in $O(\alpha_s)$ to the leading twist 2
approximation of the light-cone sum rule
for the $B\to \pi$ form factor $f^+$. Both UV and collinear 
divergences are handled by dimensional regularization
and $\overline{MS}$ renormalization. The collinear 
divergences in the hard amplitude are factorized and absorbed in the 
evolution of the light-cone wave function.
Numerically, the $O(\alpha_S)$ 
correction to the product $f_Bf^+$ amounts to 
about $30 \%$. We have shown 
that this large correction is almost completely compensated 
by the corresponding correction to the two-point sum rule for $f_B$.
The remaining $O(\alpha_s)$ effect on $f^+$ is therefore small.
This finding improves the  
accuracy and reliability of the light-cone sum rule estimate 
substantially.  

Furthermore, we have shown 
that the $O(\alpha_s)$ correction to the sum rule 
for $f_Bf^+$ depends only very weakly on  
the momentum transfer. This observation, together with the 
above-mentioned compensation strongly suggests that 
the dominant $O(\alpha_s)$ effect in the correlation 
function (\ref{corr}) comes from the 
$\gamma_5$ vertex (see Fig. 1c) which is also present 
in the two-point correlation function for $f_B$.
Thus, our calculation strongly 
supports the conjecture \cite{BKR,BBKR,BBD} that the perturbative 
correction may drop out in the ratio $f_Bf^+/f_B$.

Recently, an estimate of the perturbative correction to 
the $B\to \pi$ form factor 
was obtained \cite{Sz} in a different approach 
combining the constituent quark model for $B$ and $\pi$ 
with light-cone wave functions. Although the results agree
qualitatively it is 
difficult to directly compare our result with this 
model-dependent calculation.

A more detailed account of our calculation 
as well as applications to various 
exclusive $B$ and $D$ decays 
will be published elsewhere.

\bigskip 

{\bf Acknowledgements}.

We are grateful to A. Ali, V. Braun, A. Grozin
and A. Vainshtein for useful discussions.
This work is supported by the German Federal Ministry for 
Research and Technology (BMBF) under contract number 05 7WZ91P (0).

\newpage
%%%%%%%%%%The figures%%%%%%%%%%%%%%%

%%%%%%%%%%%%%%%%%%%%%%%%%%%%%%%%%%%%
% ffffffff---Fig.1---fffffffffffffffffffff
%\newpage
\begin{figure}
\begin{center}
\epsfig{figure=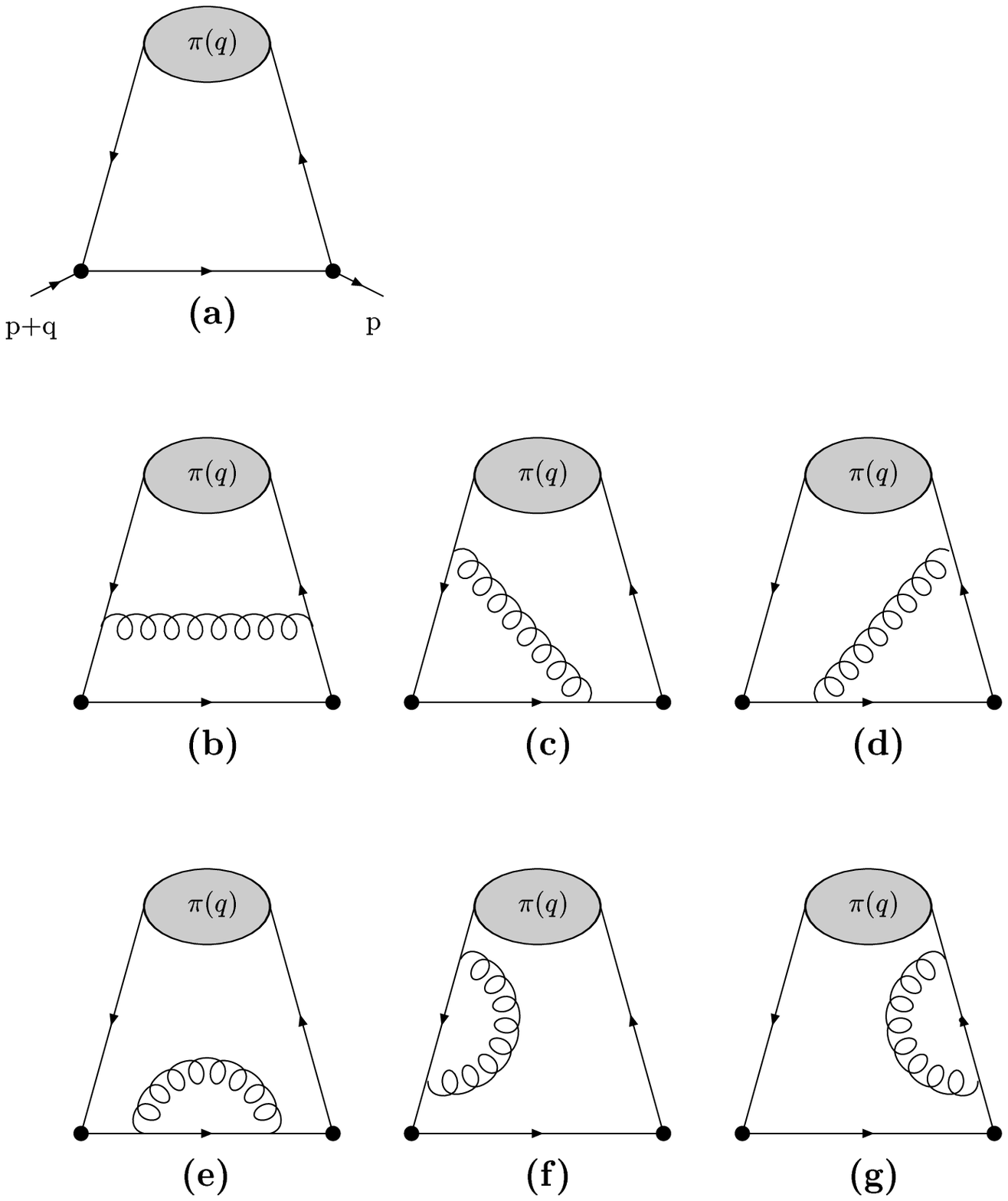}
\end{center}
\caption{Feynman diagrams contributing to the correlation
function (\ref{corr}): (a) zeroth order in 
$\alpha_s$, (b-g) first order in $\alpha_s$.}
\end{figure}
%%%%%%%%%%

% ffffffff---Fig.2---fffffffffffffffffffff
\begin{figure}
\begin{center}
\epsfig{figure=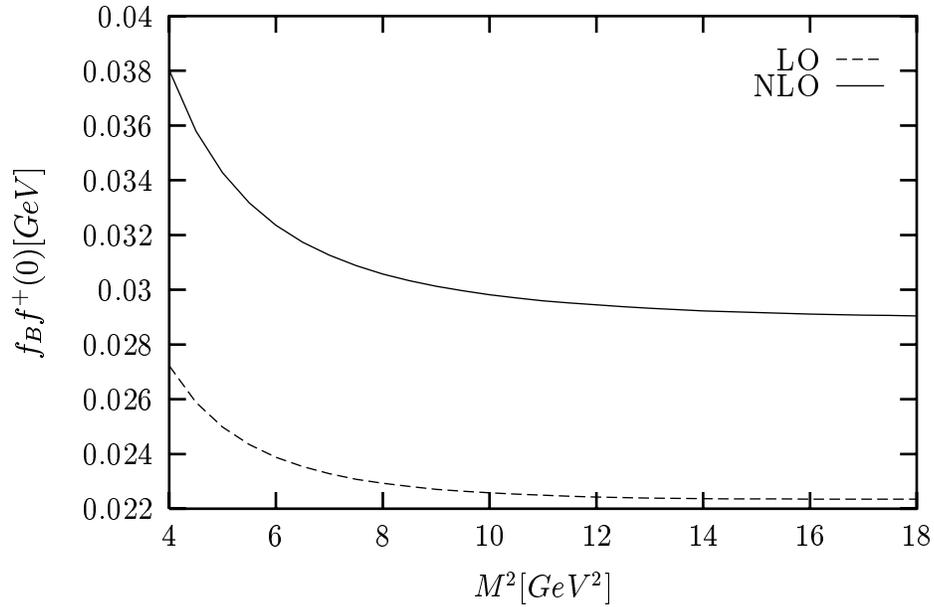}
\end{center}
\caption{Light-cone sum rule estimate for $f_Bf^+(0)$ in leading twist 2
approximation as a function of the Borel parameter $M^2$: NLO (solid ) in
comparison to LO (dashed).}
\end{figure}
%%%%%%%%%%%%%%%%

% ffffffff---Fig.3---fffffffffffffffffffff
\begin{figure}
\begin{center}
\epsfig{figure=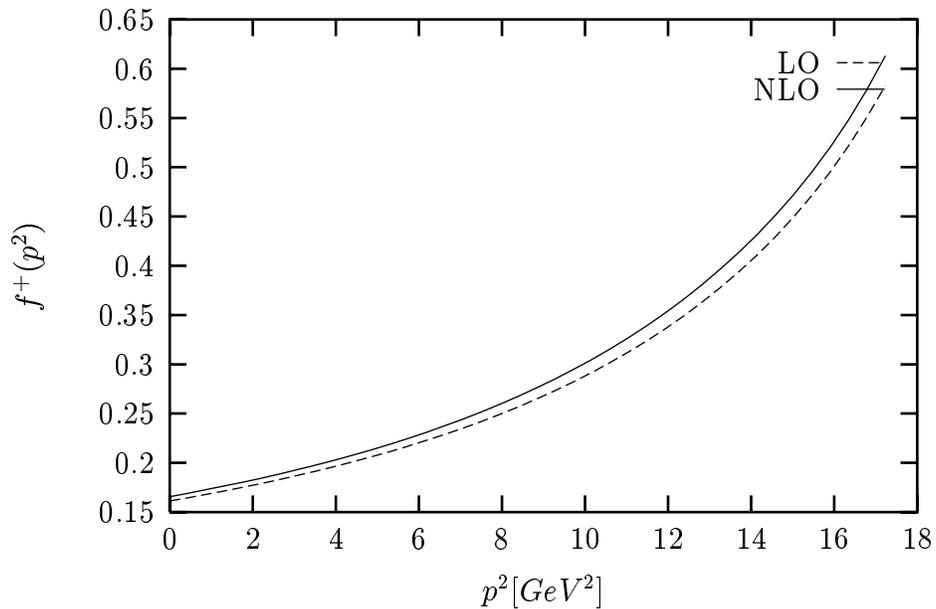}
\end{center}
\caption{Momentum dependence of the form factor $f^+(p^2)$ in leading twist 2
approximation: LO (dashed) in comparison to NLO (solid).}
\end{figure}
%%%%%%%%%%%%%%%%%%%%%%%

% ffffffff---Fig.4---fffffffffffffffffffff
\begin{figure}
\begin{center}
\epsfig{figure=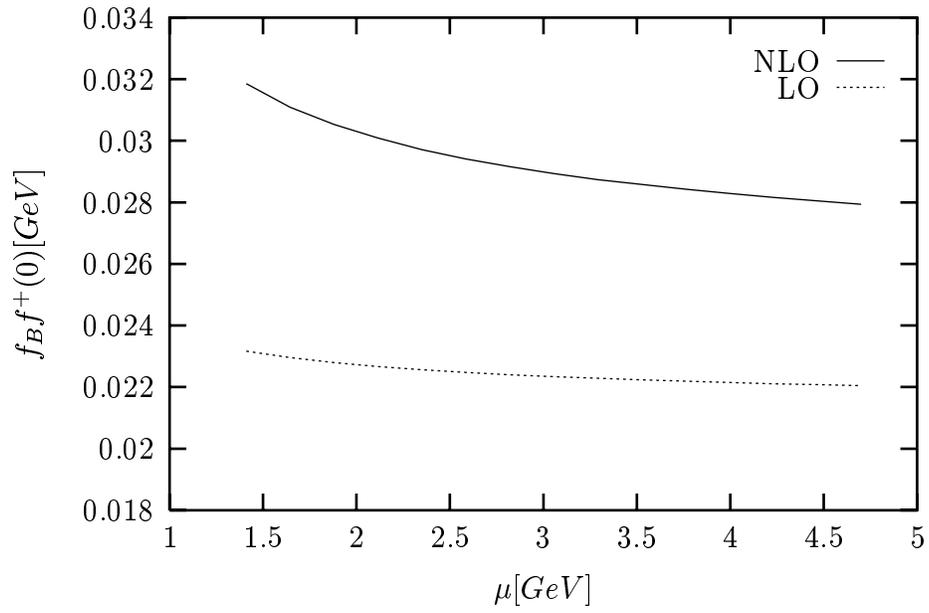}
\end{center}
\caption{Scale dependence of the light-cone sum rule estimate of 
$f_Bf^+(0)$ in leading twist 2 approximation: NLO (solid) in comparison to LO
(dotted).}
\end{figure}
%%%%%%%%%%%%%%%%%%%%%%%%%

\end{document}